\documentclass{article}
\usepackage{spconf,amsmath,graphicx}
\usepackage{multicol}
\usepackage{bm,amsmath,amssymb}
\usepackage{subfigure}

\title{Target Estimation in Colocated MIMO Radar via Matrix Completion}
\name{Shunqiao Sun, Athina P. Petropulu and Waheed U. Bajwa \thanks{This work was supported by the Office of Naval Research under Grant
ONR-N-00014-12-1-0036.}}
\address{ Department of Electrical \& Computer Engineering,
Rutgers, The State University of New Jersey}
\begin{document}
%
\maketitle
\begin{abstract}
We consider a colocated MIMO radar scenario, in which the receive antennas forward their measurements to a fusion center. Based on the received data, the fusion center formulates a matrix which is then used for target parameter estimation. When the receive antennas sample the target returns at  Nyquist rate, and assuming that there are more receive antennas than targets, the data matrix at the fusion center is low-rank.
When each receive antenna sends to the fusion center only a small number of samples, along with the sample index,  the receive data matrix has  missing elements, corresponding to the samples that were not forwarded. Under certain conditions, matrix completion techniques can be applied to recover the full receive data matrix, which can then be used in conjunction with array processing techniques, e.g., MUSIC,   to obtain target information.
Numerical results indicate that good target recovery can be achieved with occupancy of the receive data matrix as low as 50\%.

\end{abstract}
\begin{keywords}
Array processing, compressed sensing, matrix completion, MIMO radar, MUSIC
\end{keywords}
\section{Introduction}
\label{sec:intro} Multiple-input and multiple-output (MIMO) radar systems
have received considerable attention in recent years due to their superior
target estimation performance. Colocated MIMO radar systems  exploit waveform
diversity to formulate a long virtual array with number of elements equal to
the product of the number of transmit and receive antennas. As a result, they
achieve  higher resolution than traditional phased array radars for the same
amount of data \cite{Stoica: 07m}\cite{chen}. Compressed sensing (CS) enables
MIMO radar systems to maintain their advantages while  significantly reducing
the required measurements per receive antenna \cite{Herman}\cite{yao}. In
CS-based MIMO radar, target parameters are estimated by exploiting the
sparsity of targets in the angle, Doppler and range space, referred to as the
\emph{target space}. For CS-based sparse target estimation, the target space
needs to be discretized into a fine grid, based on which the CS sensing
matrix is constructed. However, performance of CS-based MIMO radar degrades
when targets fall between grid points, a case also known as basis mismatch
\cite{chi} in the CS literature.

Another approach related in spirit to CS is that of matrix completion. Matrix
completion aims to recover a low-rank data matrix from partial samples of its
entries by solving a relaxed nuclear norm optimization problem
\cite{completion1}\cite{completion2}. Array signal processing with matrix
completion has been studied in \cite{Waters}\cite{wangxin}. To the best of
our knowledge, however, matrix completion has not been exploited for target
estimation in colocated MIMO radar. Our paper is related to the ideas in
\cite{wangxin} in the sense that matrix completion is applied to the received
data matrix formed by an array. However, due to the unique structure of the
received signal in MIMO radar, the problem formulation and treatment in here
is different than that in \cite{wangxin}.

The main idea of our work is as follows. We consider a colocated MIMO radar
scenario in which receive antennas forward their measurements to a fusion
center. Based on the received data, the fusion center formulates a matrix,
which is then used for estimating the target parameters. When the receive
antennas sample the target returns at  Nyquist rate, and assuming that there
are more receive antennas than targets, the data matrix at the fusion center
is low-rank. When each receive antenna sends to the fusion center only a
small number of samples, along with the sample index,  the receive data
matrix has  missing elements, corresponding to the samples that were not
forwarded. Under certain conditions, matrix completion techniques can be
applied to recover the full receive data matrix, which can then be used in
conjunction with  parametric methods such as MUSIC to obtain target
information. Compared to CS MIMO radar, our proposed method has the same
advantage in terms of reduction of samples needed for accurate estimation but
it avoids the basis mismatch issue inherent in CS-based approach.

The rest of the paper is organized as follows. The colocated MIMO radar
system model is described in Section \ref{model}. Background of noisy
matrix completion is introduced in Section 3. The applicability of matrix
completion to MIMO radar is discussed in Section 4, and numerical results are
given  in Section 5. Finally, Section 6 provides concluding remarks.

\section{Colocated MIMO Radar System }
\label{model}

\begin{figure}
\centering
\includegraphics[width=3.4in]{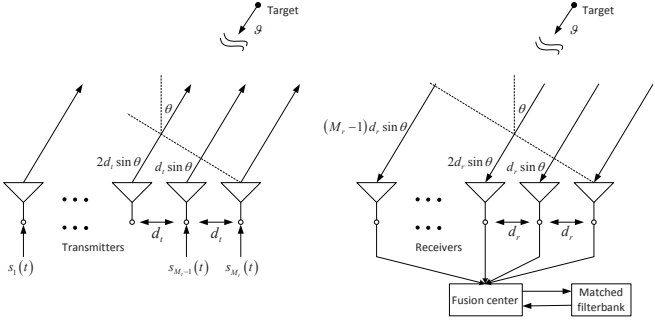}
\caption{Colocated MIMO radar system under the ULA model. }\label{MIMO_Model}  
\end{figure}
We consider a MIMO radar system that employs colocated transmit and receive
antennas, shown in Fig. \ref{MIMO_Model}. We use $M_t$ and $M_r$ to denote
the numbers of transmit antennas and receive antennas, respectively. Although
our results can be extended to an arbitrary antenna configuration, the
results here are presented for the case in which  the transmit and receive
antennas form a uniform linear array (ULA) with inter-element spacing between
transmit and receive antennas $d_t$ and $d_r$, respectively. We further
assume ${d_t} = {d_r} = {\lambda \mathord{\left/
 {\vphantom {\lambda  2}} \right.
 \kern-\nulldelimiterspace} 2}$, where $\lambda$ is the wavelength of the carrier
 signal. Further, the waveforms ${s_k}\left( t \right), k=1,...,M_t$,
 transmitted from the transmit antennas are assumed to be narrow-band and
 orthogonal.
 Now suppose there are $K$ point targets in the far field at angles $\theta_k, k=1,...,K$, each moving with speed $\vartheta_k$.
  The corresponding ${M_t} \times K$ dimensional transmit steering matrix can be expressed as ${\bf{A}}\left( {{\theta }} \right) = \left[ {{\bf{a}}\left( {{\theta _{{1}}}} \right),...,{\bf{a}}\left( {{\theta _{{K}}}} \right)} \right]$, where
 \begin{align}
{\bf{a}}\left( {{\theta _k}} \right) = {\left[ {1,{e^{j\frac{{2\pi }}{\lambda }{d_t}\sin \left( {{\theta _k}} \right)}},...,{e^{j\frac{{2\pi }}{\lambda }\left( {{M_t} - 1} \right){d_t}\sin \left( {{\theta _k}} \right)}}} \right]^T},
 \end{align}
and the ${M_r} \times K$ dimensional receive steering matrix ${\bf{B}}\left( {{\theta }} \right)$ can be defined in a similar fashion based on the receive steering vectors ${\bf b}(\theta_k)$.

In order to estimate the speed of each target, multiple pulses need to be
transmitted. Let $Q$ be the number of transmitted pulses, with $T_{PRI}$
being  the pulse repetition interval. Assume the target reflection
coefficients $\left\{ {{\beta _k}} \right\},k = 1,...,K$ are complex and
remain constant during the $Q$ pulses. For slowly moving targets, ($2
\vartheta T_p/\lambda \ll 1$, where $T_p$ is the pulse duration) the Doppler
shift within a pulse can be ignored, while the Doppler  changes from pulse to
pulse.


Under the narrowband assumption  the received signal at the $l$th receive antenna can be approximated as \cite{yao}
  \begin{equation}
{x_{l}}\left( t \right) \approx  \sum\limits_{k = 1}^K {{\beta _k}{e^{j\frac{{2\pi }}{\lambda }2{\vartheta_k}t}}{b_l}\left( {{\theta _k}} \right){{\bf{a}}^T}\left( {{\theta _{{k}}}} \right)} {\bf{s}}\left( t \right)+ {w_{l}}\left( t \right).
 \end{equation}
 Here ${\bf{s}}\left( t \right) = {\left[ {{s_1}\left( t \right),...,{s_{{M_t}}}\left( t \right)} \right]^T}$.

 Suppose that each receive antenna samples the received signal at rate ${L \mathord{\left/
 {\vphantom {L {{T_s}}}} \right.
 \kern-\nulldelimiterspace} {{T_s}}}$ and forwards the samples  to the fusion center. Here, $T_s$ is the sampling time, $L$ is the number of nonzero samples and we assume $L \gg
 K$.
At the fusion center, received data during the $q$th pulse can be written as
\cite{parafac2}
\begin{align} \label{data_matrix}
{{\bf{X}}_q} = {\bf{B}}\left( {{\theta }} \right){\bf{\Sigma }}{{\bf{D}}_q}{{\bf{A}}^T}\left( {{\theta }} \right){\bf{S}} + {{\bf{W}}_q}  = {{\bf{Z}}_q} + {{\bf{W}}_q},
\end{align}
where ${\bf{\Sigma }} = {\rm{diag}}\left( {\left[ {{\beta _1},...,{\beta _K}} \right]} \right)$; ${{\bf{D}}_q} = {\rm{diag}}\left( {{{\bf{d}}_q}} \right)$, with ${{\bf{d}}_q} = {\left[ {{e^{j2\pi 2{\vartheta_1}\left( {q - 1} \right){T_{PRI}}}},...,{e^{j2\pi 2{\vartheta_K}\left( {q - 1} \right){T_{PRI}}}}} \right]^T}$;
 ${\bf{S}} = \left[ {{\bf{s}}\left( {0{T_s}} \right),...,{\bf{s}}\left( {\left( {L - 1} \right){T_s}} \right)} \right] $; and ${{\bf{W}}_q} $ is a Gaussian noise matrix. 
Note that both matrices ${\bf{\Sigma }}$ and ${{\bf{D}}_q}$ are rank-$K$,
while the rank of matrix $\bf S$ is $\min \left\{ {{M_t},L} \right\}$. Thus,
for ${M_t} > K$ the rank of the noise free data matrix ${{\bf{Z}}_q}
={\bf{B}}\left( {{\theta }} \right){\bf{\Sigma
}}{{\bf{D}}_q}{{\bf{A}}^T}\left( {{\theta }} \right){\bf{S}}$ is $K$. In
other words, the data matrix ${{\bf{Z}}_q}$ is low-rank based on the
assumption that ${M_r} \gg K$.

\section{Matrix Completion with Noise}
We now provide a brief overview of the problem of recovering a rank $r$
matrix ${\bf M} \in {{\mathbb{C}}^{{n_1} \times {n_2}}}$ based on partial
knowledge of its entries, possibly corrupted by noise, i.e., ${\left[
{\bf{Y}} \right]_{ij}} = {\left[ {\bf{M}} \right]_{ij}} + {\left[ {\bf{E}}
\right]_{ij}}, \left( {i,j} \right) \in \Omega$, where, ${\left[ {\bf{E}}
\right]_{ij}}$ represents noise and  $\Omega$ is the set of observed entries.
This can also be expressed as ${{\mathcal P}_\Omega }\left( {\bf{Y}} \right)
= {{\mathcal P}_\Omega }\left( {\bf{M}} \right) + {{\mathcal P}_\Omega
}\left( {\bf{E}} \right)$, where ${{\mathcal P}_\Omega }$ represents  the
sampling operation.
 According to \cite{completion2}, when $\bf M$ is low-rank and its singular vectors are sufficiently spread, i.e., both the numbers of zero and large elements in the singular vectors are not large,  $\bf M$ can be recovered by solving  a relaxed nuclear norm optimization problem,
 given by
\begin{align} \label{robust_mc}
 \min \;\;&{\left\| {\bf{X}} \right\|_*} \quad \quad
 {\rm{s.t.}}\;\; &{\left\| {{{\mathcal P}_\Omega }\left( {{\bf{X - Y}}} \right)} \right\|_F} \le \delta
 \end{align}
where ${\left\|  \cdot  \right\|_*}$ denotes the nuclear norm, i.e., the sum of singular values of $\bf X$, while $\delta > 0$ is a constant. 

To test the `sufficiently spread' requirement, the \emph{strong incoherence
property} of matrix $\bf M$ with parameter $\mu $ has been introduced in
\cite{completion1}. Consider the singular value decomposition (SVD) of $\bf
M$, i.e., ${\bf{M}} = \sum\limits_{k = 1}^r {{\rho
_k}{{\bf{u}}_k}{\bf{v}}_k^H}$, and define ${P_U} = \sum\limits_{1 \le i \le
r} {{{\bf{u}}_i}{\bf{u}}_i^H}$, ${P_V} = \sum\limits_{1 \le i \le r}
{{{\bf{v}}_i}{\bf{v}}_i^H}$,
and  ${\bf{T}} = \sum\limits_{1 \le i \le r} {{{\bf{u}}_i}{\bf{v}}_i^H}$.
When the following conditions are satisfied, the matrix $\bf M$ is said to
satisfy the strong incoherence property with $\mu  = \max \left( {{\mu
_1},{\mu _2}} \right)$.

{\bf{A1}}) For all pairs $\left( {a,a'} \right) \in \left[ {{n_1}} \right] \times \left[ {{n_1}} \right]$ and $\left( {b,b'} \right)\in \left[ {{n_2}} \right] \times \left[ {{n_2}} \right]$, there is ${\mu _1} > 0$ such that
\begin{align}
 \left| {\left\langle {{{\bf e}_a},{P_U}{{\bf e}_{a'}}} \right\rangle  - \frac{r}{{{n_1}}}{1_{a = a'}}} \right|& \le {\mu _1}\frac{{\sqrt r }}{{{n_1}}}, \label{condition1} \\
 \left| {\left\langle {{{\bf e}_b},{P_V}{{\bf e}_{b'}}} \right\rangle  - \frac{r}{{{n_2}}}{1_{b = b'}}} \right|& \le {\mu _1}\frac{{\sqrt r }}{{{n_2}}}, \label{condition2}
 \end{align}
where ${{\bf e}_a}$ is the vector with the $a$th element equal to $1$ and others being zero, while $1_{a = a'}$ indicates that it is equal to $1$ when ${a = a'}$ and $0$ otherwise.

{\bf{A2}}) For all $\left( {a,b} \right) \in \left[ {{n_1}} \right] \times \left[ {{n_2}} \right]$, there exists a constant ${\mu _2} > 0$ such that $\left| {{{{T}}_{ab}}} \right| \le {\mu _2}\frac{{\sqrt r }}{{\sqrt {{n_1}{n_2}} }},
$
 where ${ {T}}_{ab}$ is the $\left( {a,b} \right)$ entry of the matrix $\bf T$.

Note that it has been shown in \cite{completion1} that when the maximum (in
magnitude) values of the $K$ left and right singular vectors are bounded,
i.e.,
\begin{align}
{\left\| {{{\bf u}_k}} \right\|_{{\ell_\infty }}} \le \sqrt {\frac{{{\mu _B}}}{{{n_1}}}}, {\left\| {{{\bf v}_k}} \right\|_{{\ell_\infty }}} \le \sqrt {\frac{{{\mu _B}}}{{{n_2}}}}
\end{align}
with ${\mu _B} = O\left( 1 \right)$, then the strong incoherence property is
guaranteed with $\mu  \le {\mu _B}\sqrt r $.

Now define $n= \max \left( {{n_1},{n _2}} \right)$ and suppose $\bf M$
satisfies the strong incoherence property. Then \cite{completion1}
establishes in the noiseless case that, by observing $N$ randomly selected
entries with $N \ge C{\mu ^2}nr{\log ^6}n$ for some constant $C$, the matrix
$\bf M$ can be recovered exactly with a probability of at least $1 - {n^{ -
3}}$. Further, \cite{completion2} establishes that, when observations are
corrupted with white noise ${\left[ {\bf{E}} \right]_{ij}}$ that is zero-mean
Gaussian with variance ${\sigma ^2}$, the recovery error is bounded as
${\left\| {{\bf{M - \hat M}}} \right\|_F} \le 4\sqrt {\frac{1}{p}{\left( {2 +
p} \right)\min \left( {{n_1},{n_2}} \right)}} \delta + 2\delta$,
  where $p = \frac{N}{{{n_1}{n_2}}}$ is the fraction of observed entries.

\section{Matrix Completion for MIMO Radar}

The left singular vectors of ${\bf Z}_q$ defined in (\ref{data_matrix}) are the eigenvectors of ${{\mathbf{Z}}_q}{\mathbf{Z}}_q^H  = {\mathbf{HS}}{{\mathbf{S}}^H}{{\mathbf{H}}^H}$, where ${\bf H}={\bf{B}}\left( {{\theta }} \right){\bf{\Sigma }}{{\bf{D}}_q}{{\bf{A}}^T}\left( {{\theta }} \right)$. The right singular vectors of ${\bf Z}_q$ are the eigenvectors of ${{\bf{S}}^H}{{\bf{H}}^H}{\bf{HS}}$. Since $\bf S$ is orthogonal, it holds that ${\mathbf{S}}{{\mathbf{S}}^H} = {\mathbf{I}}$. Thus, the left singular vectors are only determined by  matrix ${\bf H}$, while the right singular vectors are affected by both transmit waveforms and matrix ${\bf H}$.

For the problem considered in this paper, it is difficult to determine
analytically the behavior of the entries of the left and right singular
vectors of ${\bf Z}_q$. Instead, we get an idea of the behavior of the
maximum values and the parameters that affect their spread using simulations.
We consider a MIMO radar setup in which
 the target direction of arrival (DOA) angles are uniformly distributed in $\left[ { - {{90}^ \circ },{{90}^ \circ }} \right]$ and the corresponding target speeds are uniformly distributed in the range $\left[ {150,450} \right]{m \mathord{\left/ {\vphantom {m s}}
 \right.
 \kern-\nulldelimiterspace} s}$. In addition, $\beta_k$ are following complex Gaussian distribution and kept unchanged for $Q=10$ pulses. The pulse repetition interval is ${T_{PRI}} = {1 \mathord{\left/
 {\vphantom {1 {4000}}} \right.
 \kern-\nulldelimiterspace} {4000}}$ second, and the carrier frequency is $f= {10^9}$ Hz, resulting in $\lambda  = {c \mathord{\left/
 {\vphantom {c f}} \right.
 \kern-\nulldelimiterspace} f} = 0.3$ meter. Two types of orthogonal waveforms are considered: Hadamard and Gaussian orthogonal waveforms. Several cases of parameters are verified. Case I: $M_r=40$, $L=128$; Case II: $M_r=1000$, $L=128$; Case III: $M_r=40$, $L=1024$. Each case runs for 300 iterations.

\begin{figure}[t]
\centering
\subfigure[]{\includegraphics[width=1.6in]{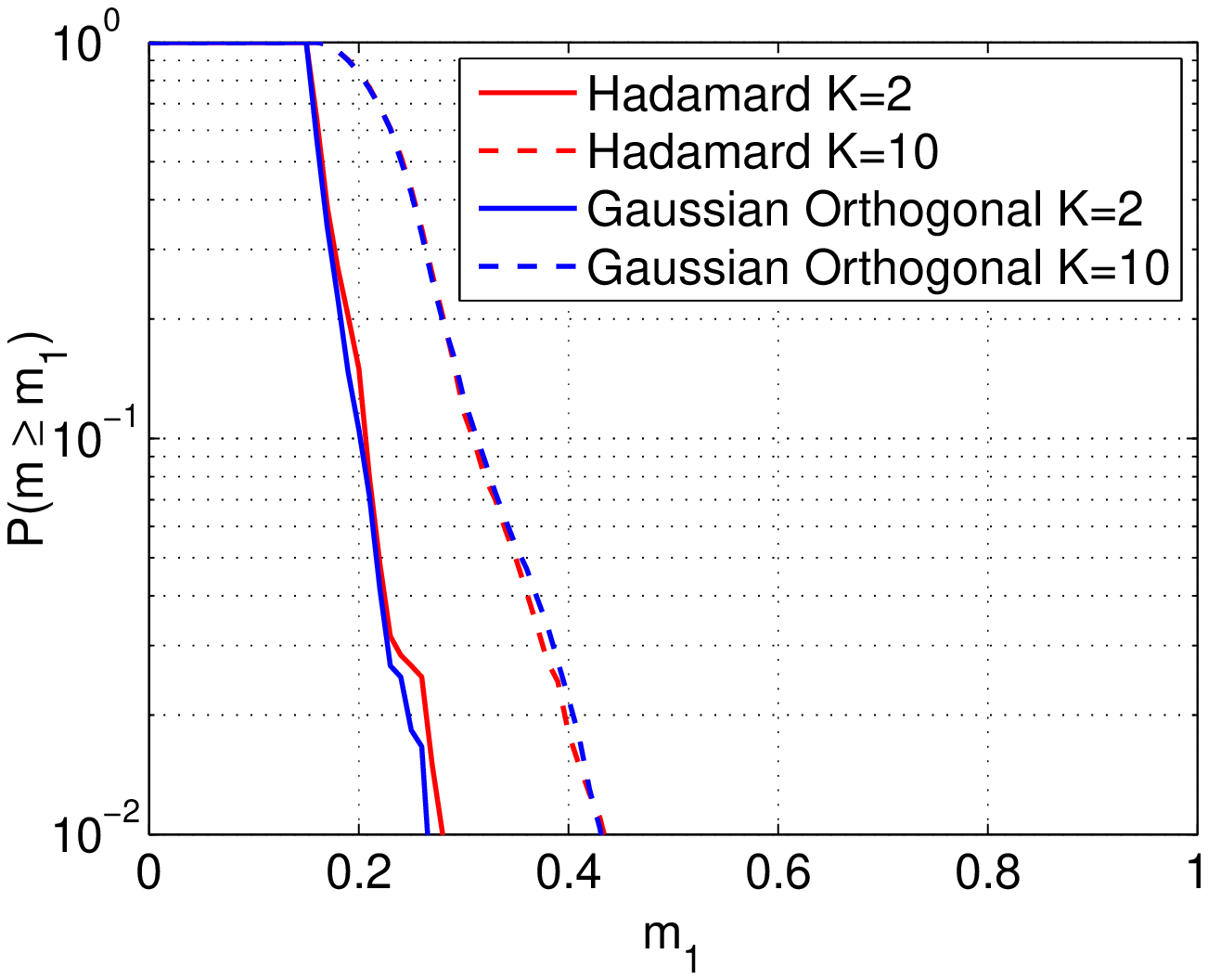}}
\subfigure[]{\includegraphics[width=1.6in]{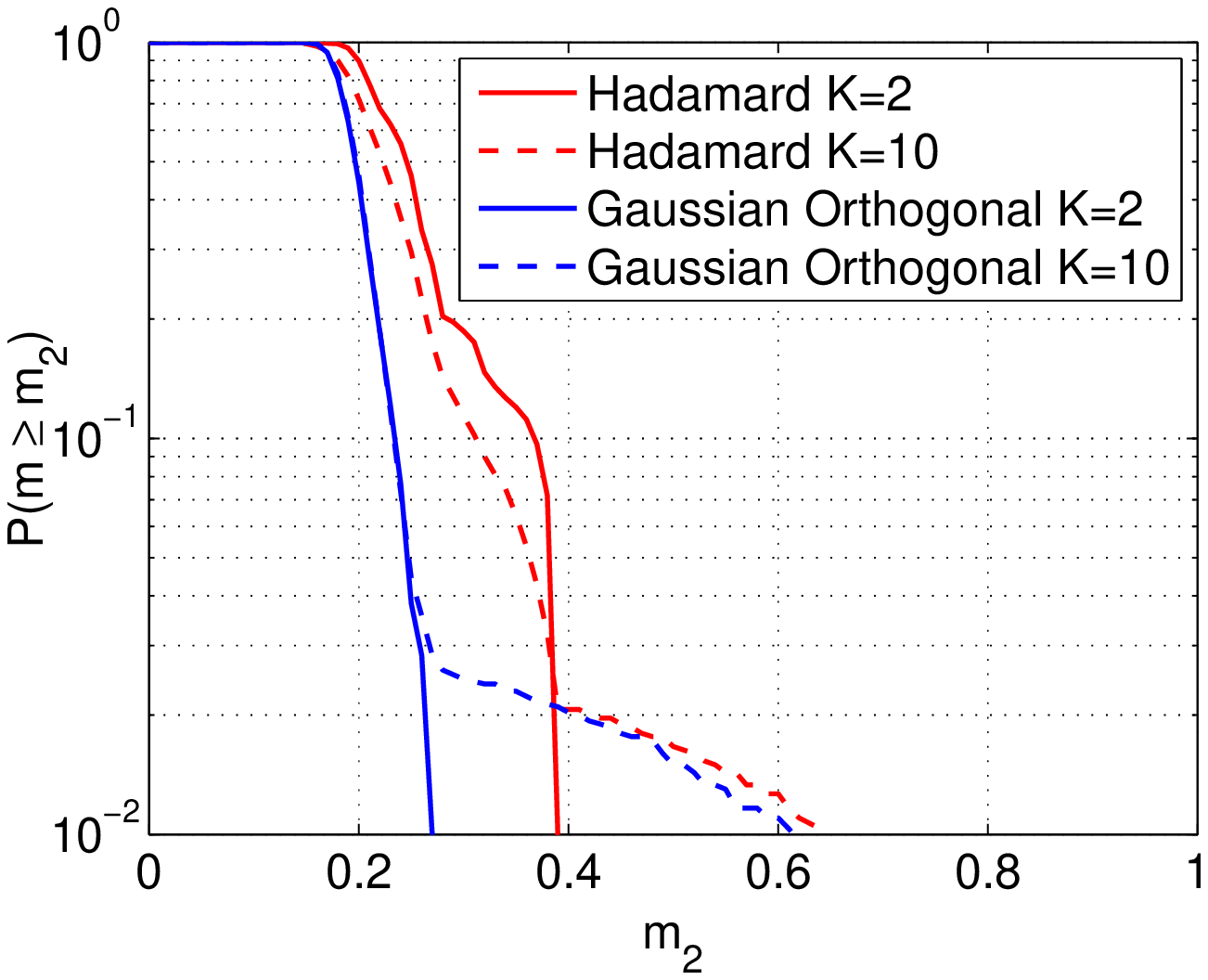}}
\subfigure[]{\includegraphics[width=1.6in]{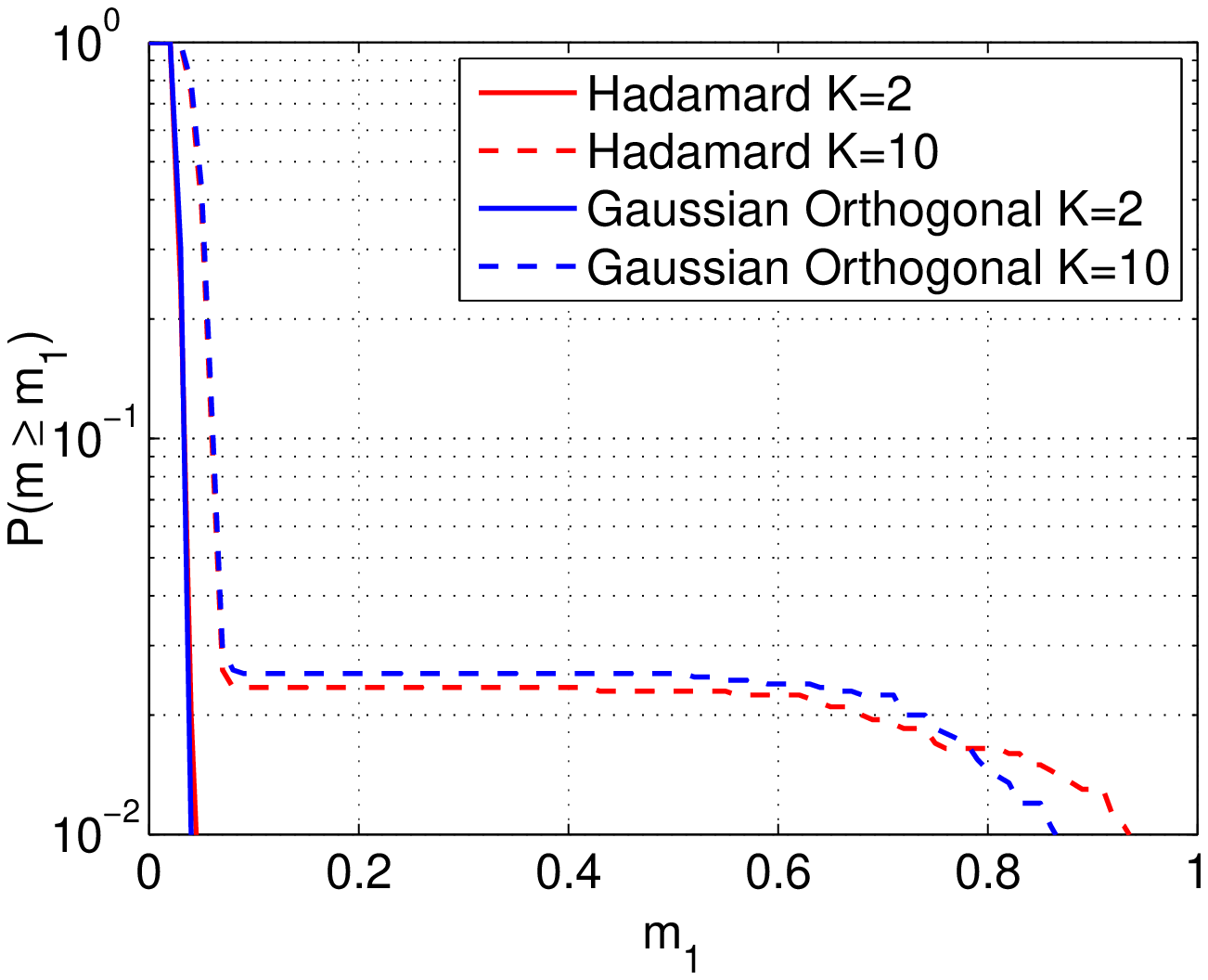}}
\subfigure[]{\includegraphics[width=1.6in]{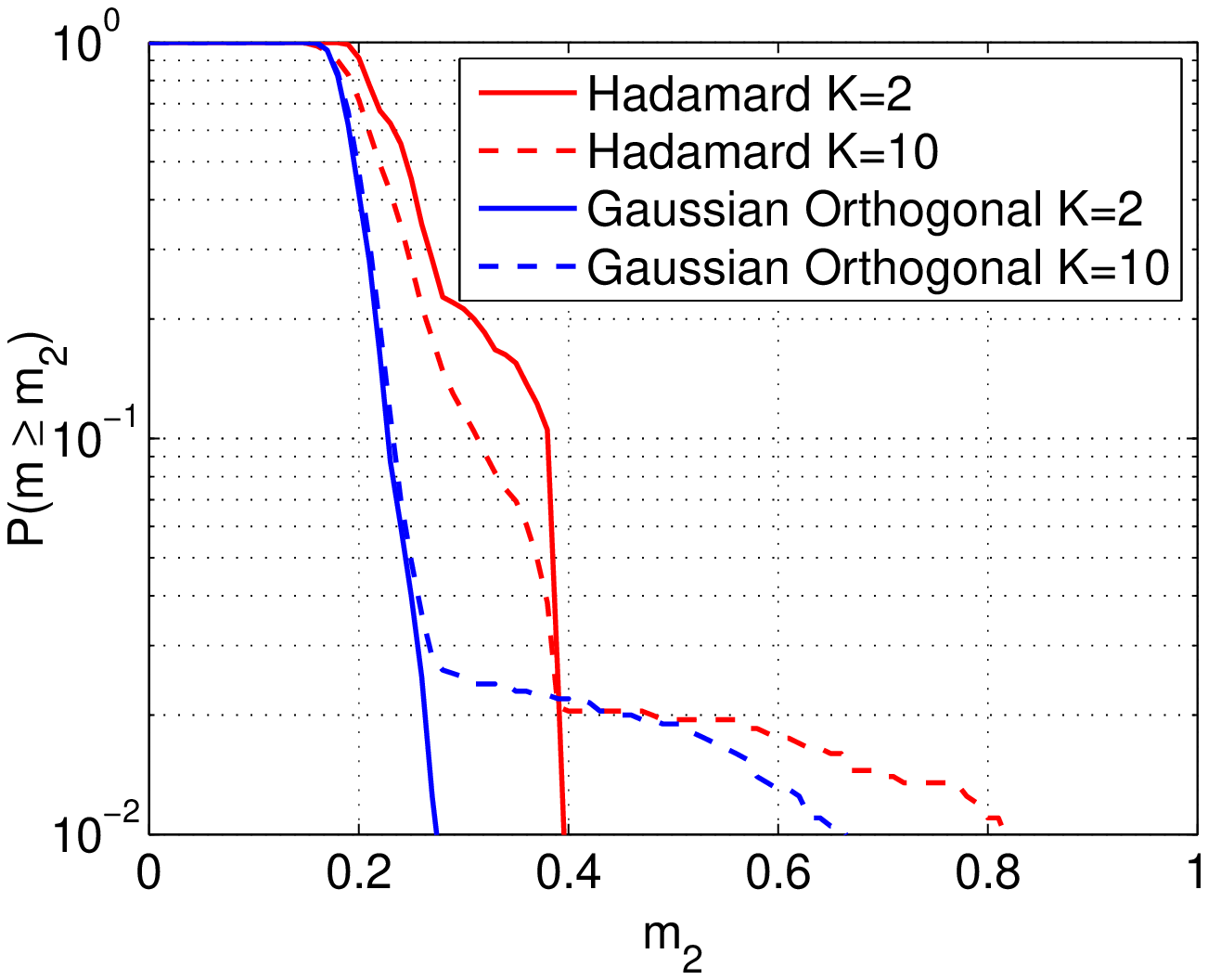}}
\caption{ Complementary CDF of $m_1$ and $m_2$. Case I:  $M_r=40$, $L=128$;  (a) left singular vectors, (b) right singular vectors; Case II:  $M_r=1000$, $L=128$; (c) left singular vectors, (d) right singular vectors.}  \label{ICDF}
\end{figure}

 Let $m_1$ and $m_2$ denote the maximum element value of $K$ left and right singular vectors of ${\bf Z}_q$. The complementary cumulative distribution function (CDF) curves of $m_1$ and $m_2$, i.e., $\Pr \left( {m > {m_i}} \right), i=1,2$ are plotted in Fig. \ref{ICDF} for Cases I and II. 
It can be seen from these plots that as $M_r$ increases, the bounds of $m_1$ for both $K=2$ and $K=10$ decrease, while the distribution of $m_2$ does not
significantly change when $L$ is fixed. Figure \ref{ICDF} (c) shows that as $M_r$ gets large, $m_1$ gets bounded by a small number with high probability.  
Space limitations prevent us from displaying more figures of Case III, which
show that $m_2$ is bounded by a small number with high probability as $L$
gets large.

\begin{figure}[t]
\centering
\subfigure[]{\includegraphics[width=1.6in]{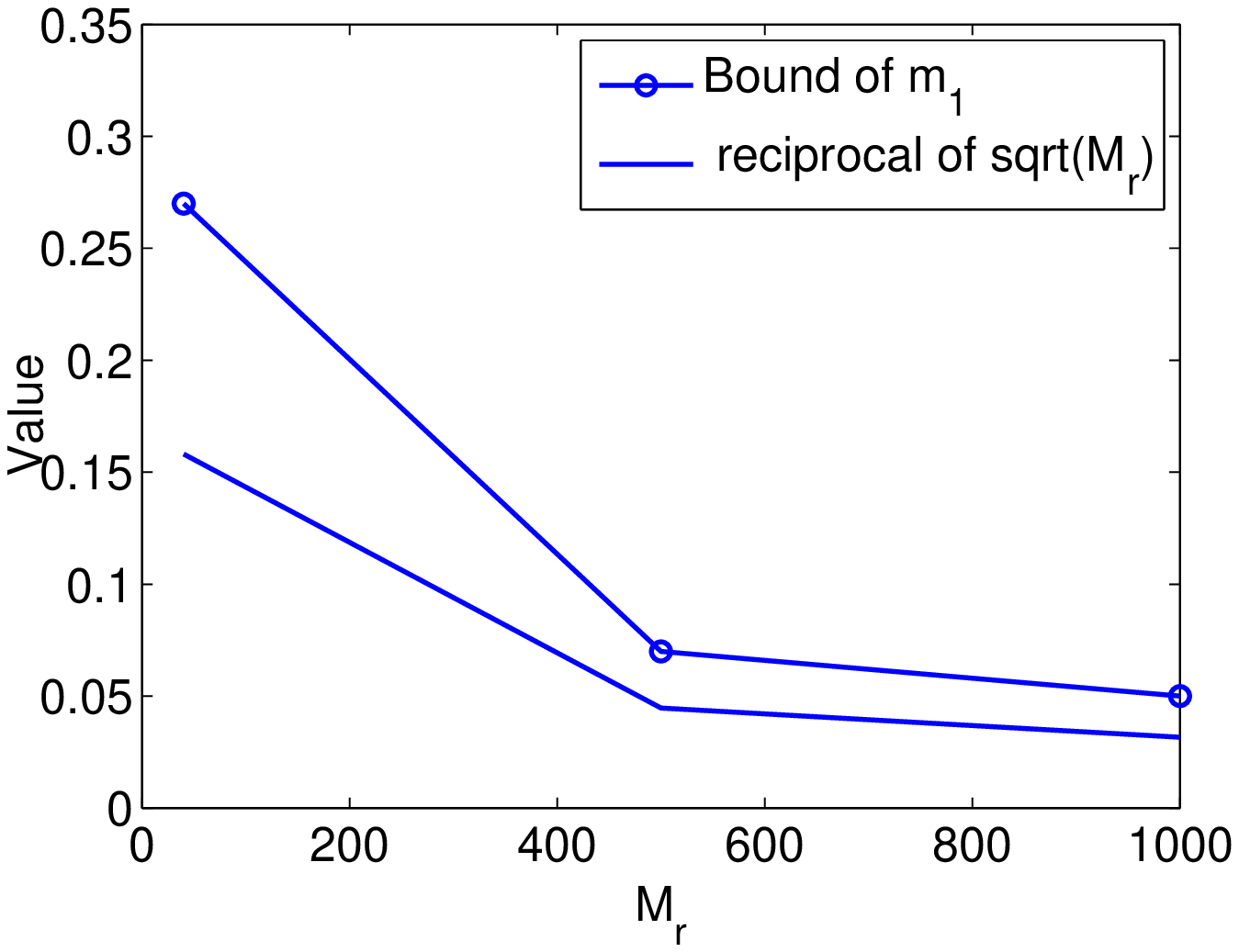}}
\subfigure[]{\includegraphics[width=1.6in]{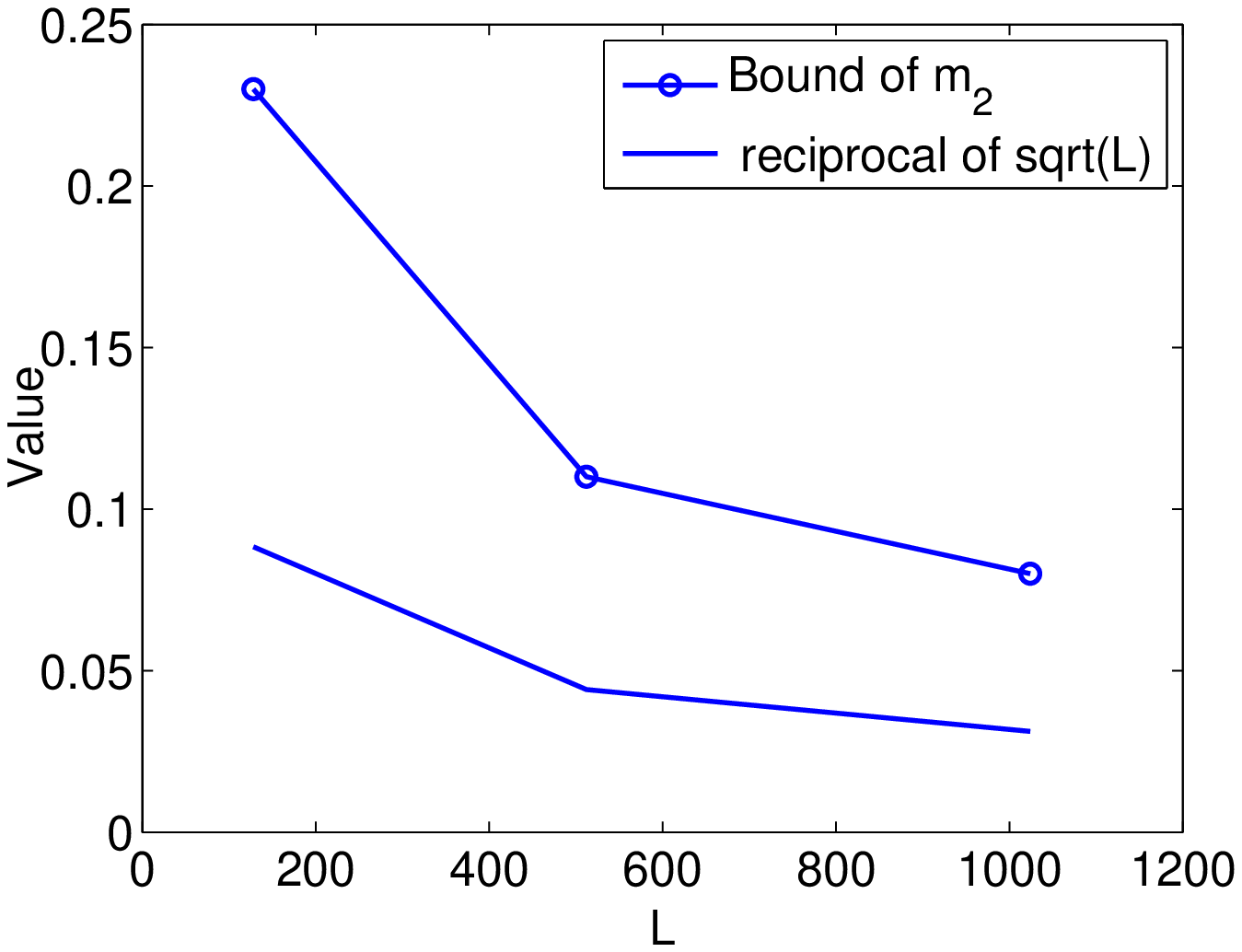}}
\caption{Bounds of $m_1$ and $m_2$ scales with reciprocal of ${\sqrt {{M_r}} }$ and ${\sqrt {{L}} }$, respectively.} \label{scaling}
\end{figure}

Extensive simulations also show that the bounds on $m_1$ and $m_2$ scale as
${1 \mathord{\left/ {\vphantom {1 {\sqrt {{M_r}} }}} \right.
 \kern-\nulldelimiterspace} {\sqrt {{M_r}} }}$ and ${1 \mathord{\left/
 {\vphantom {1 {\sqrt {{L}} }}} \right.
 \kern-\nulldelimiterspace} {\sqrt {{L}} }}$, respectively, with some constant $\sqrt {{\mu _B}} $.  For Gaussian orthogonal waveform and $K=2$, ${\mu _B} \approx 2.4$ for the bound of $m_1$ (see Fig. \ref{scaling} (a)) and ${\mu _B} \approx 6.5$  for the bound of $m_2$ (see Fig. \ref{scaling} (b)). For $K=10$ and Hadamard waveform, the scaling laws also hold but with larger constants $\mu_B$. Therefore, depending on the number of receive antennas $M_r$ and the number of samples $L$ in one pulse, $m_1$ and $m_2$ can be assumed to be bounded by small numbers.  Based on \cite{completion1}, therefore, we conclude that the strong incoherence property
 is likely
 satisfied in our problem.

It is also worth noting that under Gaussian orthogonal waveforms, $m_2$  is concentrated in a smaller range as compared with the range for Hadamard waveforms. This indicates that the  waveform indeed plays a role for the use of matrix completion, and perhaps the waveform can be optimally designed to result in low probability of large values for $m_2$. 



Since the matrix completion conditions appear to be satisfied in our case, we
propose that each antenna obtains and forwards to the fusion center a small
number of samples during each pulse. Note that along with the samples, the
antenna needs to inform the fusion center on how the sample was obtained so
that the fusion center can determine where to position the received samples
in the receive data matrix ${\bf Z}_q$. In a practical setting, a
pseudo-random sampling ADC at each receive antenna could be used in the place
of the Nyquist sampler, where the pseudo-random generator seeds would be
distributed to the receive antennas by the fusion center. Finally, the fusion
center can recover the data matrix ${{\bf{
 \hat
 X}}_q}$ by applying matrix completion to the received data.

Once the data matrix ${\bf \hat X}_q$ is recovered, it can go through a
matched-filter bank to produce
\begin{align}
{{\bf{Y}}_q} = \frac{1}{L}{{\bf{ \hat X}}_q}{{\bf{S}}^H} = {\bf{B}}\left( {{\theta }} \right){\bf{\Sigma }}{{\bf{D}}_q}{{\bf{A}}^T}\left( {{\theta }} \right) + {{{\bf{\tilde W}}}_q},
\end{align}
where ${{{\bf{\tilde W}}}_q}$ is noise  whose distribution is a function of
the additive noise ${\bf W}_q$ and the nuclear norm minimization problem in
(\ref{robust_mc}). Next, stacking  ${{\bf{Y}}_q} \in {{\mathbb C}^{{M_r}
\times {M_t}}}$ into a vector  ${{\bf{y}}_q} $, and based on the vectors
corresponding to
 $Q$ pulses, the following matrix can be formed: ${{\bf{Y}}_R} = \left[ {{{\bf{y}}_1},...,{{\bf{y}}_Q}} \right]  \in {{\mathbb C}^{{M_r}{M_t} \times Q}}$.
Reshaping  ${\bf Y}_R$ into ${\bf{Y}} \in {{\mathbb{C}}^{Q{M_t} \times {M_r}}}$,  we have
\begin{align}
{\bf{Y}} ={\bf{F}}{\bf{\Sigma }}\left[ {{\bf{b}}\left( {{\theta _1}} \right),...,{\bf{b}}\left( {{\theta _K}} \right)} \right] + {\bf{W}},
\end{align}
where ${\bf{F}}=\left[ {{\bf{d}}\left( {{\vartheta_1}} \right) \otimes
{\bf{a}}\left( {{\theta _1}} \right),...,{\bf{d}}\left( {{\vartheta_K}}
\right) \otimes {\bf{a}}\left( {{\theta _K}} \right)} \right]$,
${\bf{d}}\left( \vartheta \right) = {\left[ {1,{e^{j2\pi
2\vartheta{T_{PRI}}}},...,{e^{j2\pi 2\vartheta\left( {Q - 1}
\right){T_{PRI}}}}} \right]^T}$, with $ \otimes $ denoting the Kronecker
product. The sampled covariance matrix of the receive data signal can then be
obtained as ${{\bf{ \hat R}}_{Y}} = \frac{1}{M_r}{\bf{Y}}{{\bf{Y}}^H}$, based
on which target estimation can be implemented using any array processing
method such as MUSIC.


\section{Numerical Results}
In this section we present some simulation results on the performance of the proposed method.
 We use the simulation setting considered in the previous section, i.e., ${M_t} = 20, \ {M_r} = 40, \ Q=5$, $L=128$. The signal-to-noise ratio (SNR) is set to $25$ dB, while $\beta_k$ are following complex Gaussian distribution and kept unchanged for $Q$ pulses.  
   For  matrix completion, the TFOCS software package \cite{TFOCS} is used.

\begin{figure}[t]
\centering
\subfigure[]{\includegraphics[width=1.6in]{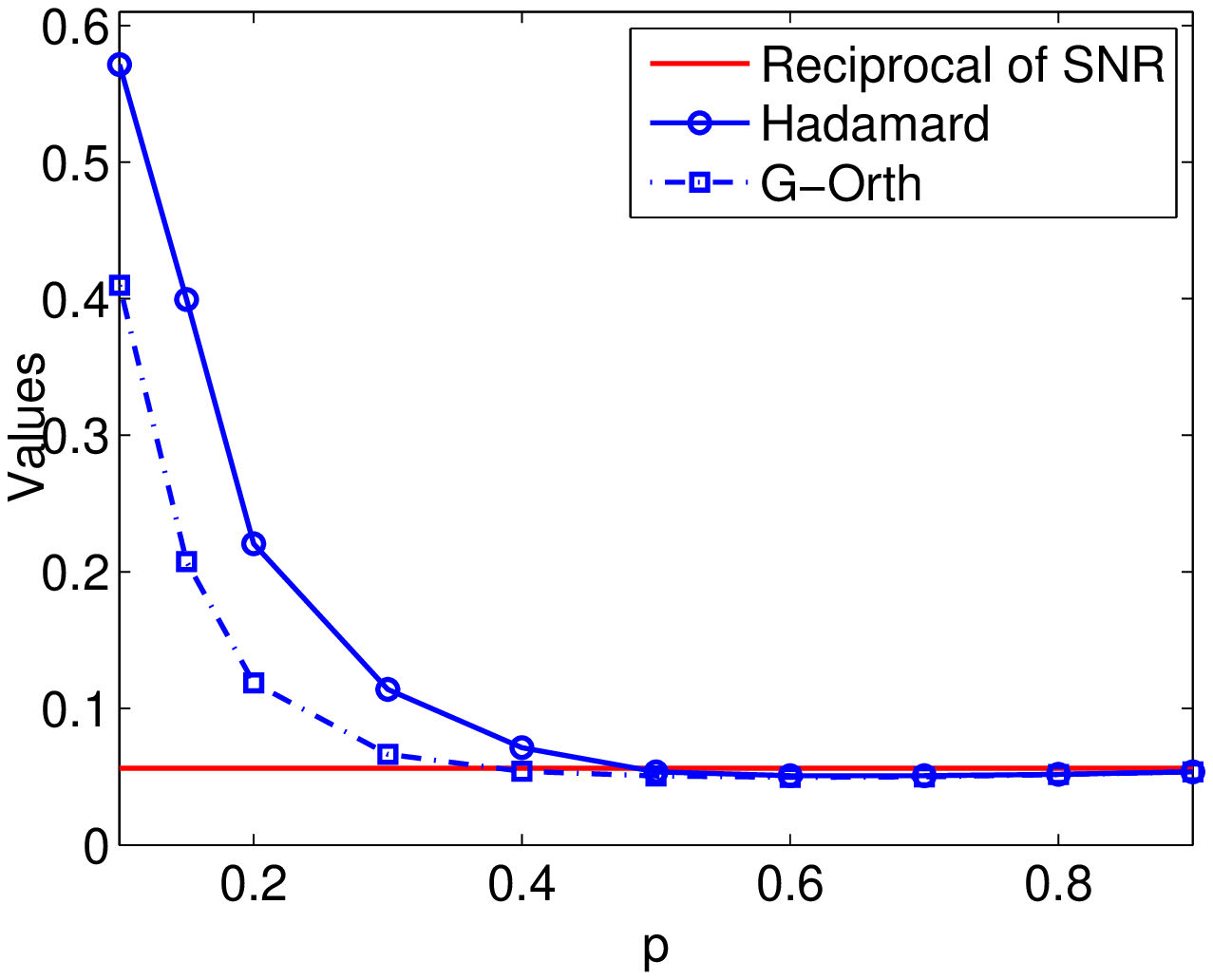}}
\subfigure[]{\includegraphics[width=1.6in]{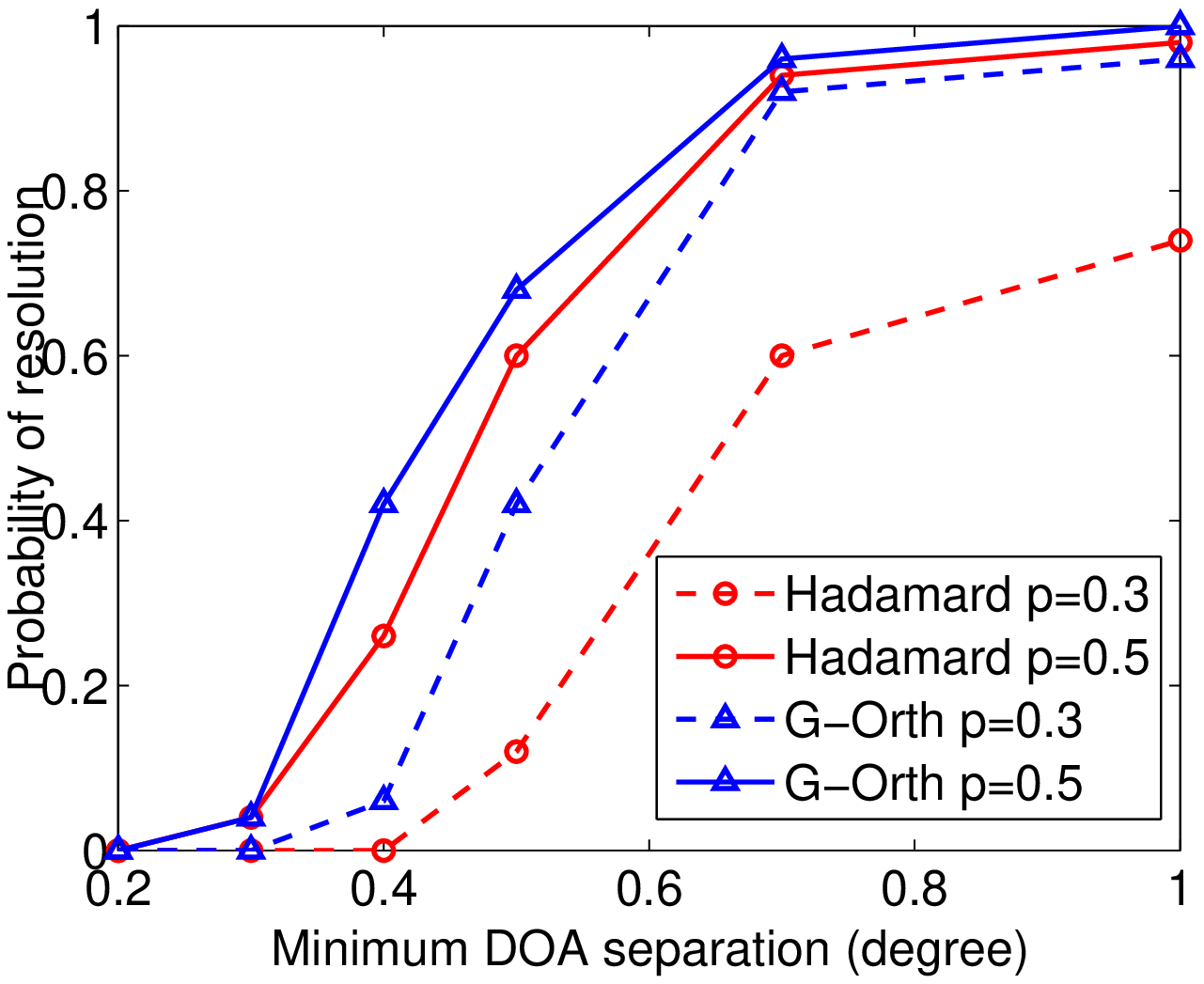}}
\caption{Performance comparisons: (a) Relative error of the recovered data matrix; (b) Probability of DOA estimation resolution.} \label{comp}               
\end{figure}

First, we plot relative errors (averaged over $50$ Monte Carlo runs) of the
received data matrix ${\bf \hat X}_q$ for Hadamard and Gaussian orthogonal
(G-Orth) waveforms. The relative error is defined as $\frac{{{{\left\|
{{\bf{Z}}_q - {\bf{\hat X}}}_q \right\|}_F}}}{{{{\left\| {\bf{Z}}_q
\right\|}_F}}}$, where ${\bf Z}_q$ is the data matrix calculated without
missing elements. Under each waveform,
$K=2$ point targets in the far field are randomly generated. 
The result is shown in Fig. \ref{comp} (a) for $q=1$. It can be seen from
this figure that, as $p$ increases, the relative recovery error of the data
matrix under Gaussian orthogonal waveform reduces to the reciprocal of the
SNR faster than that under Hadmard waveform. A plausible reason for this is
that under the Gaussian orthogonal waveform, the maximum value of elements in
the singular vectors of ${\bf Z}_q$ is bounded by a smaller number with high
probability, as compared with that under the Hadamard waveform (see Fig.
\ref{ICDF}).

Next, the probabilities of DOA estimation resolution under the two orthogonal
waveforms are plotted in Fig. \ref{comp} (b) for the following scenario. Two
targets are randomly generated among DOA range $[-20^\circ, 20^\circ]$ with
minimum DOA separations ${d_\theta } = \left[ {{{0.2}^ \circ },{{0.3}^ \circ
},{{0.4}^ \circ },{{0.5}^ \circ },{{0.7}^ \circ },{1^ \circ }} \right]$ and
the corresponding speeds are set to $150$ and $400$ ${m \mathord{\left/
 {\vphantom {m s}} \right.
 \kern-\nulldelimiterspace} s}$. The MUSIC algorithm is applied to obtain the target DOA information. If the DOA estimates ${\hat \theta }_i$, $i=1,2$ satisfy $\left| {\theta_i  - {\hat \theta}_i } \right| \le \varepsilon {d_\theta }, \varepsilon  = 0.1$, we declare this as a success. The probability of DOA resolution is then defined as the fraction of successful events in $50$ iterations.
It can be seen from the figure that when $p=0.3$, the Gaussian orthogonal
waveform has a much better DOA estimation resolution compared with Hadamard
waveform. As $p$ increases to $0.5$, the performance difference becomes small
since the relative recovery errors under both waveforms are similar (see Fig.
\ref{comp} (a)). Figure \ref{comp} confirms that Gaussian orthogonal
waveforms are better than Hadamard waveforms for matrix completion-based DOA
estimation.

%
%

\section{Conclusions}
We have provided results suggesting that matrix completion can be used in MIMO radar to
  reduce the number of data needed to be communicated to the fusion center by each receive antenna.
  Numerical results show that   matrix completion in conjunction with MUSIC  can achieve accurate target estimation with sub-Nyquist samples. Thus, the proposed method can result in significant savings in terms of data that need to be obtained at the receive antennas and subsequently transmitted to the fusion center.



\end{document}